\documentclass[runningheads,a4paper]{article}

\usepackage{epsfig}
\usepackage{amssymb}
\usepackage{url}
\usepackage{amsmath}
\usepackage{latexsym}
\usepackage{multirow}

\long\def\comment#1{}

\newtheorem{definition}{Definition}
 \newtheorem{theorem}{Theorem}
   
 \newtheorem{example}{Example}  
 \newtheorem{remark}{Remark}  
   
 \newtheorem{proposition}{Proposition}    
 \newtheorem{corollary}{Corollary}

 \long\def\comment#1{}

\newcommand{\naturals}{\mathbb{N}}
\newcommand{\positivenaturals}{\mathbb{N}_{>0}}
\newcommand{\integers}{\mathbb{Z}}

\newcommand{\rationals}{\mathbb{Q}}

\newcommand{\reals}{\mathbb{R}}
\newcommand{\nonnegativereals}{\mathbb{R}_0}

\renewcommand{\emptyset}{\varnothing}
\renewcommand{\phi}{\varphi}

\newcommand{\pr}[1]{\mbox{\tt #1}}   

\def\defemb#1#2{\expandafter\def\csname #1\endcsname
							  {\relax\ifmmode #2\else\hbox{$#2$}\fi}}
\defemb{cA}{{\cal A}}
\defemb{cB}{{\cal B}}
\defemb{cC}{{\cal C}}
\defemb{cD}{{\cal D}}
\defemb{cE}{{\cal E}}
\defemb{cF}{{\cal F}}
\defemb{cG}{{\cal G}}
\defemb{cH}{{\cal H}}
\defemb{cI}{{\cal I}}
\defemb{cJ}{{\cal J}}
\defemb{cK}{{\cal K}}
\defemb{cL}{{\cal L}}
\defemb{cM}{{\cal M}}
\defemb{cN}{{\cal N}}
\defemb{cO}{{\cal O}}
\defemb{cP}{{\cal P}}
\defemb{cQ}{{\cal Q}}
\defemb{cR}{{\cal R}}
\defemb{cS}{{\cal S}}
\defemb{cT}{{\cal T}}
\defemb{cU}{{\cal U}}
\defemb{cV}{{\cal V}}
\defemb{cX}{{\cal X}}
\defemb{cY}{{\cal Y}}
\defemb{cZ}{{\cal Z}}

\defemb{mP}{{\sf P}}

\long\def\comment#1{}

\makeatletter
\newenvironment{prog}{\vspace{0.7ex}\par
\setlength{\parindent}{0.7cm}
\obeylines\@vobeyspaces\tt}{\vspace{0.7ex}\noindent
}
\makeatother
\newcommand{\startprog}{\begin{prog}}
\newcommand{\stopprog}{\end{prog}\noindent}

\makeatletter
\newenvironment{smallprog}{\vspace{0.7ex}\par
\setlength{\parindent}{0.7cm}
\obeylines\@vobeyspaces\tt\small}{\vspace{0.7ex}\noindent
}
\makeatother
\newcommand{\fstartprog}{\begin{smallprog}}
\newcommand{\fstopprog}{\end{smallprog}\noindent}

\makeatletter
\newenvironment{nismallprog}{\vspace{0.7ex}\par
\setlength{\parindent}{0.0cm}
\obeylines\@vobeyspaces\tt\small}{\vspace{0.7ex}\noindent
}
\makeatother
\newcommand{\fnistartprog}{\begin{nismallprog}}
\newcommand{\fnistopprog}{\end{nismallprog}\noindent}

\newcommand{\muterm}{\mbox{\sc mu-term\/}}

\newcommand{\Jambox}{{\sf Jambox}}

\newcommand{\Symbols}{{\cF}}

\newcommand{\Variables}{{\cX}}
\newcommand{\TermsOn}[2]{{\cT(#1,#2)}}

\newcommand{\Terms}{{\TermsOn{\Symbols}{\Variables}}}

\newcommand{\Var}{{\cal V}ar} 
\newcommand{\var}{{\cV}ar}

\newcommand{\toStarPosSub}[2]{{\;\mbox{$\stackrel{#1}{\longrightarrow}\hspace{.1cm}\hspace{-.2cm}^*_{#2}\,$}}}
\newcommand{\toPlusPosSub}[2]{{\;\mbox{$\stackrel{#1}{\longrightarrow}\hspace{.1cm}\hspace{-.2cm}^+_{#2}\,$}}}

\newcommand{\activationlazyrew}[1]{\stackrel{\sf A}{\to}}
\newcommand{\activationlazyrewp}[1]{\toPlusPosSub{\sf A}{}}
\newcommand{\activationlazyrews}[1]{{\toStarPosSub{\sf A}{}}}

\newcommand{\SemDomain}{{\sf A}}

\newcommand{\compOp}{\bowtie}
\newcommand{\genconstraint}{\sf c}

\newcommand{\cmatrix}[2]{{\bf #2}_{#1}}

\newcommand{\diag}{{\sf diag}}

\newcommand{\blockvectors}[3]{T_{#1,#2}(#3)}

\begin{document}


  \title{From matrix interpretations over the rationals to matrix interpretations over the naturals\thanks{Partially supported by EU (FEDER) and MICINN grant TIN 2007-68093-C02-02.
  The final publication of this paper is
   available at {\tt www.springerlink.com}}}


\author{
  Salvador Lucas\\
  ELP Group, DSIC, Universidad
  Polit\'ecnica de Valencia\\
  Camino de Vera s/n, 46022 Valencia, Spain
}

\date{}

\maketitle

\begin{abstract}
%
Matrix interpretations generalize linear polynomial interpretations
and have been proved useful in the implementation of tools for automatically proving termination
of Term Rewriting Systems.
In view of the successful use of rational coefficients in polynomial interpretations, 
we have recently generalized traditional matrix interpretations (using \emph{natural} 
numbers in the matrix entries) to incorporate \emph{real} numbers.
However, existing results 
which formally prove that polynomials over the reals are \emph{more powerful} than
polynomials over the naturals for proving termination of rewrite systems \emph{failed} to be extended
to matrix interpretations.
In this paper we get deeper into this problem.
We show that, under some conditions, it is possible to transform a matrix interpretation over the rationals satisfying a set
of symbolic constraints into a
matrix interpretation over the naturals (using \emph{bigger} matrices)
which still satisfies the constraints.
%
\\

\end{abstract}

\section{Introduction}

Constraint solving is an essential technique for the implementation of automatic verification
systems.
Many verification problems can be expressed as sets of symbolic constraints which have
to be tested for satisfaction or even solved to give some explicit solution certifying their satisfaction.
For instance,  termination problems are often
expressed as conjunctions of \emph{weak} or \emph{strict}
\emph{symbolic constraints} like  $e\succeq e'$ or $e\succ e'$
between expressions $e$ and $e'$
coming from (parts of) the programs \cite{BorRub_OrderingsAndConstraintsForTermination_RCP07}.
%
%
%
Automatic termination tools 
have to \emph{check} these constraints
and eventually provide an appropriate certificate.
A standard approach is using algebraic interpretations
which translate the \emph{symbolic} constraints into some kind of \emph{arithmetic}
constraints. 
\begin{example}\label{ExPaperRationalMatrices}
Consider the following Term Rewriting System (TRS) $\cR$ \cite{AlaLucNav_ProvingTerminationMatrixInterpretationsOverReals_WST09,%
AlaLucNav_UsingMatrixInterpretationsOverReals_PROLE09}:
{\footnotesize
\begin{eqnarray}
f(f(X)) & \to & f(g(f(X)))\label{EmbeddingRule}\\
f(g(f(X))) & \to & X\label{ProjectionRule}
\end{eqnarray}}%
A proof of termination with \emph{dependency pairs} can be easily obtained as 
follows \cite{ArtGie_TermOfTermRewUsingDepPairs_TCS00}:
Consider the following rules (called \emph{dependency pairs}) associated to $\cR$: 
%
\begin{eqnarray}
F(f(X)) & \to & F(g(f(X)))\label{EmbeddingDP1}\\
F(f(X)) & \to & F(X)\label{EmbeddingDP2}
\end{eqnarray}
The following polynomial intepretation with \emph{rational} coefficients 
%
\[
\begin{array}{rcl@{\hspace{1cm}}rcl@{\hspace{1cm}}rcl}
{}[f](x) &=& 2x+2 & [g](x) &=& \frac{1}{2}x+\frac{1}{2} & 
{}[F](x) &=& x
\end{array}
\]
can be used to prove termination of $\cR$ by showing that $[l]\geq[r]$ for the rules 
$(\ref{EmbeddingRule}),(\ref{ProjectionRule})$ (where $[l]$ and $[r]$ are the interpretations of
terms $l$ and $r$ which is obtained by structural induction), and
$[u]>[v]$ for the dependency pairs $(\ref{EmbeddingDP1}),(\ref{EmbeddingDP2})$.
\end{example}
A recent and fruitful approach is using \emph{matrix interpretations} \cite{EndWalZan_MatrixInterpretations_JAR08}, 
where the $k$-ary symbols 
$f$ are given \emph{parametric} matrix functions $[f]$, e.g.,
$F_1x_1+\cdots+F_kx_k+F_0$, where the $F_i$'s are (square) matrices 
of some fixed dimension $n$ and $F_0$ is an $n$-tuple.
The variables $x_1,\ldots,x_k$ are intended to range on $n$-tuples as well.
In \cite{EndWalZan_MatrixInterpretations_JAR08}, only natural numbers are used both in matrices and $n$-tuples.

\begin{example}\label{ExPaperRationalMatricesWithMatrices}
The following matrix interpretation \emph{over the naturals}
\[\begin{array}{rcl@{\hspace{1cm}}rcl@{\hspace{1cm}}rcl}
{}[f](x) & = & {} 
\left (
\begin{array}{cc}
1 & 1 \\
1 & 1
\end{array} \right ) 
x + \left (
\begin{array}{c}
1\\
1
\end{array} \right ) &
{}[g](x) & = & {} \left (
\begin{array}{cc}
0 & 1 \\
0 & 0
\end{array} \right ) x \\[0.5cm] 
{}[F](x) & = & {} \left (
\begin{array}{cc}
1 & 1 \\
0 & 1
\end{array} \right ) x
\end{array}
\]
can also be used for proving termination of $\cR$ in Example \ref{ExPaperRationalMatrices}.
Both interpretations have been automatically obtained by using \muterm\ \cite{Lucas_MUTERM_RTA04}.
\end{example}
In \cite{AlaLucNav_ProvingTerminationMatrixInterpretationsOverReals_WST09,%
AlaLucNav_UsingMatrixInterpretationsOverReals_PROLE09}, 
Endrullis et al.'s framework was extended to matrices containing \emph{real numbers} in the
entries.
The adaptation was motivated by a number of recent theoretical works and
experimental evaluations showing that \emph{polynomial} interpretations over the 
rationals can be advantageously used instead of  polynomial interpretations over the 
naturals 
\cite{BorrallerasEtAl_SolvingPolyAritithSMT_CADE09,FuhsEtAl_SearchTechRatPolyOrd_AISC08,%
Lucas_RelPowPoly_AAECC06}.
%
%
In \cite{FuhsEtAl_SearchTechRatPolyOrd_AISC08}, syntactic conditions ensuring that, when dealing with \emph{linear} polynomial interpretations, 
real coefficients \emph{must be} used for addressing the corresponding termination problem, were given for the first time.
The extension of these results to 
matrix interpretations over the reals failed \cite{AlaLucNav_ProvingTerminationMatrixInterpretationsOverReals_WST09,%
AlaLucNav_UsingMatrixInterpretationsOverReals_PROLE09}.

Thus, the following question arises: Are rational numbers somehow \emph{unnecessary} when dealing with matrix interpretations?
By examining the proofs of termination contributed to the International Competition on Termination\footnote{See \url{http://termcomp.uibk.ac.at/termcomp/}}
by tools like \Jambox\ 
\cite{Endrullis_Jambox}, which makes extensive use of matrix interpretations over the naturals, 
and comparing them to the corresponding ones generated by tools like \muterm, which
emphasizes the use of polynomials over the rationals \cite{Lucas_PolOverRealsTheoPrac_TIA05}, 
one may notice that, in many cases, proofs of termination with polynomials over the rationals 
somehow correspond to proofs using matrix intepretations whose matrices have specific shapes.
Examples \ref{ExPaperRationalMatrices} and \ref{ExPaperRationalMatricesWithMatrices}
illustrate this connection, which we substantiate in this paper: the bigger are the \emph{values} of the (possibly rational) coefficients in the polynomial interpretation 
the more non-null entries are in the corresponding matrix coefficients.
In this paper we investigate this phenomenon.
In Section \ref{SecNumbersAsMatrices} we develop a notion of
\emph{numeric matrix representation} which permits a representation of a natural number as
a matrix of (smaller) natural numbers. This representation preserves the usual arithmetics of
natural numbers (addition and product).
Therefore, we can think of such matrices as having a \emph{value}; the value of the number from
which they were obtained.
Then, in Section \ref{SecTransformingMatrices} we show how to extend this process to
transform a matrix of natural numbers into a (bigger) matrix of (smaller) natural numbers. 
As a consequence, we prove that every matrix of natural numbers can be represented as a
\emph{bit} matrix with entries in $\{0,1\}$ which still preserves its `value' and arithmetic behavior.
In Section \ref{SecRepOfRationalNumbers}, we address the problem of representing 
arbitrary \emph{rational numbers} as matrices of integer numbers.
We argue that this is possible only for finite subsets of rational numbers.
We investigate the use of \emph{nilpotent} matrices (i.e., square matrices which become null after a
finite number of selfproducts) as suitable devices for achieving this.
In Section \ref{SecMatrixBlockBasedMatrixInterp}, we introduce a new generalization of 
matrix interpretations, which we call \emph{block-based matrix interpretations}.
Essentially, we view a matrix as structured into blocks of (sums of)
\emph{constant} or \emph{scalar} matrices.
%
%
In Section \ref{SecTransfConstraints}, we investigate how the satisfaction of different kind of constraints
is \emph{preserved} under the matrix transformations investigated in the previous sections.
As a consequence of our results, we prove that TRSs which can be proved terminating by using
matrix interpretations over the naturals can also be proved terminating by using a matrix interpretation
based on \emph{bit} matrices. Furthermore, we show that, under some conditions, proofs of termination
which are carried out by polynomial or matrix intepretations over the rationals can also be obtained
by using matrix interpretations over the naturals (like in Example \ref{ExPaperRationalMatricesWithMatrices}).
Section \ref{SecConclusions} summarizes our contribution and concludes.


\section{Numbers as matrices}\label{SecNumbersAsMatrices}

In the following, we use the standard notations and terminology for matrices 
\cite{LanTis_TheoryOfMatrices_1985,Meyer_MatrixAnalysisAppliedAlgebra_2000,%
Zhang_MatrixTheory_1999}.
Given $p,q\in\positivenaturals$ and a set of numbers $N$ (usually $\reals$, $\rationals$, or
$\naturals$), we write $A\in N^{p\times q}$ to say that $A$ is a matrix of $p$ rows and $q$ columns
with entries $A_{ij}\in N$ (a $p\times q$-matrix for short). If $p=q$, then $A$ is a \emph{square} matrix.

We investigate the representation of (rational) numbers
as matrices of integer numbers satisfying some structural properties.
%
Of course, we want to ensure that the representation preserves (part of) the algebraic structure
of the considered numeric domain, in such a way that, for instance, the arithmetic operations
and orderings among numbers (of the considered kind) can be implemented by using 
matrix  operations and orderings.
%
%
%

We view a rational number as a product 
$p\frac{1}{q}$ for an integer number $p\in\integers$ and 
a positive natural number $q\in\positivenaturals$.
First, we formalize a generic framework for representing real numbers as matrices:
%
mappings $\mu:\reals\to\reals^{m\times n}$ and 
$\rho:\reals^{m\times n}\to\reals$ 
provide a representation of real numbers as matrices and
vice versa.

\begin{definition}[Numeric matrix representation]\label{DefNumMatrixRep}
Let $N\subseteq\reals$ be a subset of real numbers, $p,q\in\positivenaturals$, and 
$M\subseteq\reals^{p\times q}$ be a 
subset of matrices. A  \emph{numeric matrix representation} is a pair 
$(\mu,\rho)$ of mappings $\mu:N\to M$ 
(called a \emph{numeric representation})
and $\rho:M\to N$ (called a \emph{matrix valuation})
such that $\rho\circ\mu=id$.
\end{definition}
Let $\cmatrix{p\times q}{1}$ be the $p\times q$-matrix all whose entries contain $1$.
A matrix $C=c\cmatrix{p\times q}{1}$ for some $c\in\reals$ (i.e.,  whose entries are settled to $c$) 
is called a \emph{constant} matrix.
The identity (square) matrix of size $n$ is denoted $I_n$.
A matrix $S=cI_n$ for some $c\in\reals$ 
is called a \emph{scalar} matrix.
We consider the following numeric matrix representation.

\begin{definition}\label{DefNumericMatrixInterpReals}
Let $m,p,q\in\positivenaturals$ and $A\in\reals^{p\times q}$. We 
let 
\begin{enumerate}
\item $\mu^{p\times q}_m(x)=\frac{mx}{pq}\cmatrix{p\times q}{1}$, i.e., each real number $x$ is mapped to
a \emph{constant} $p\times q$-matrix with entries $\frac{mx}{pq}$.
\item $\rho_m(A)=\frac{\sum^p_{i=1}\sum^q_{j=1}A_{ij}}{m}$, i.e., each matrix $A$ is mapped to the
number which is obtained by adding all entries in $A$ and then dividing this number by $m$.
\end{enumerate}
\end{definition}
%
%
%
In the following, we prove some properties of $\rho_m$ which are used below. 
%
%

\begin{proposition}\label{PropAdditivityOfRepMapping}
Let $m,p,q\in\positivenaturals$ and $A,B\in\reals^{p\times q}$.
Then, $\rho_m(A + B)=\rho_m(A)+\rho_m(B)$ and
$\rho_m(\alpha A)=\alpha\rho_m(A)$ for all $\alpha\in\reals$.
\end{proposition}

\begin{proposition}\label{PropProductOfRepMapping}
Let $p,q,r\in\positivenaturals$, $A\in\reals^{p\times q}$ and $B\in\reals^{q\times r}$.
If $B$ (resp.\ $A$) is scalar and $q\leq r$ (resp.\ $q\leq p$), or $B$ (resp. $A$) is a constant matrix,
then $\rho_p(AB)=\rho_p(A)\rho_q(B)$. 
\end{proposition}
Propositions \ref{PropAdditivityOfRepMapping} and \ref{PropProductOfRepMapping} 
entail the following.

\begin{corollary}\label{CoroAdditionAndProductOfRepMapping}
Let $p,q,r\in\positivenaturals$, $A\in\reals^{p\times q}$ and $B\in\reals^{q\times r}$.
If $B$ (resp.\ $A$) is an additive combination of scalar and constant matrices,
then $\rho_p(AB)=\rho_p(A)\rho_q(B)$. 
\end{corollary}

\begin{corollary}\label{CoroProductOfRepMappingInSquareMatrices}
Let $n\in\positivenaturals$ and $A,B$ be $n$-square matrices.
If $B$ is an additive combination of scalar or constant matrices,
then $\rho_n(AB)=\rho_n(A)\rho_n(B)=\rho_n(BA)$.
\end{corollary}
%
If we consider only $n$-square matrices for representations, then $\mu'_n(x)=xI_n$ could also
be used with $\rho_n$ as a numeric matrix representation.

\begin{remark}[Use of vectors]
Since vectors $\vec{v}$ can be seen as special matrices $\vec{v}\in\reals^{n\times 1}$ of $n$ rows and a single
column, the numeric matrix representation in Definition \ref{DefNumericMatrixInterpReals} can also be used to represent real
numbers as vectors.
In particular, we get $\mu^{n\times 1}_n(x)=x\cmatrix{n}{1}$ and 
$\rho_n(\vec{v})=\frac{\sum_{i=1}^nv_i}{n}$. 
\end{remark}


\section{Transforming matrices of numbers}\label{SecTransformingMatrices}

We can \emph{extend} any numeric representation $\mu:\reals\to\reals^{p\times q}$ to a mapping 
$\mu:\reals^{m\times n}\to\reals^{m p\times nq}$ from $m\times n$-matrices $A$ into 
$mp\times nq$-matrices $\mu(A)$ by just replacing the numeric entries $A_{ij}$ in $A$ 
by the corresponding  matrices $\mu(A_{ij})$, i.e., $\mu(A)=(\mu(A_{ij}))_{i=1,j=1}^{i=m,j=n}$.
The new matrix can be viewed as a \emph{block} matrix whose blocks are $\mu(A_{ij})$ for
$1\leq i\leq m$ and $1\leq j\leq n$.

\begin{example}\label{ExTransformMatricesByEntries}
We can transform the matrix 
{\footnotesize 
$\left (
\begin{array}{cc}
3 & 0 \\
0 & 3
\end{array} \right )$}
by using 
$\mu^{p\times q}_m$ in Definition \ref{DefNumericMatrixInterpReals}; with $m=p=q=3$, we obtain:
\[\mu^{3\times 3}_3(
\left (
\begin{array}{cc}
3 & 0 \\
0 & 3
\end{array} \right )
)=
\left (
\begin{array}{cc}
\mu^{3\times 3}_3(3) & \mu^{3\times 3}_3(0) \\
\mu^{3\times 3}_3(0) & \mu^{3\times 3}_3(3)
\end{array} \right )
=
\left (
\begin{array}{cc}
\cmatrix{3\times 3}{1}& \cmatrix{3\times 3}{0} \\
\cmatrix{3\times 3}{0} & \cmatrix{3\times 3}{1}
\end{array} \right )
\]
\end{example}
The interesting feature of the matrix obtained in Example \ref{ExTransformMatricesByEntries} is 
that it is a \emph{bit} matrix whose entries are either\footnote{Matrices with entries in 
$\{0,1\}$ are called $(0,1)$-matrices in \cite[Section 8.2]{Zhang_MatrixTheory_1999}.} 
$0$ or $1$. Note that this is due to the use of $\mu^{3\times 3}_3$
which permits a representation of `$3$' as a constant matrix $\cmatrix{3\times 3}{1}$ with $1$'s only.
For the numeric matrix representation in Definition \ref{DefNumericMatrixInterpReals}, we have the following.

\begin{proposition}\label{PropValuePreservingTransf}
Let $m,n,p,q,r,s\in\positivenaturals$ and $A\in\reals^{r\times s}$.
Then, $\rho_{mn}(\mu^{p\times q}_m(A))=\rho_n(A)$.
\end{proposition}

%
\begin{theorem}\label{TheoCombinationsOfTransformedMatrices}
Let $m,n,p,q,r,s,t,u\in\positivenaturals$.
If $A,B\in\reals^{r\times s}$, then, 
$\rho_m(A)+\rho_m(B)=\rho_{mn}(\mu^{p\times q}_n(A) + \mu^{p\times q}_n(B))$ and
$\rho_m(\alpha A)= \alpha\rho_{mn}(\mu^{p\times q}_n(A))$ for all $\alpha\in\reals$.
If $A\in\reals^{s\times t}$ and $B\in\reals^{t\times u}$, then 
$\rho_m(AB)=\rho_{mn}(\mu^{p\times q}_n(A)\mu^{q\times r}_n(B))$.
\end{theorem}
Propositions \ref{PropAdditivityOfRepMapping} and \ref{PropProductOfRepMapping} 
entail the following.

\begin{corollary}\label{CoroAdditionAndProductOfTransformedMatrices}
Let $m,n,p,q,r,s\in\positivenaturals$, $A\in\reals^{m\times m}$ and $B\in\reals^{m\times s}$.
If  $A$ is an additive combination of scalar and constant matrices,
then $\rho_m(AB)=\rho_{mn}(\mu^{p\times q}_n(A)\mu^{q\times r}_n(B))=\rho_{mn}(\mu^{p\times q}_n(A))\rho_{mn}(\mu^{q\times r}_n(B))$.
\end{corollary}
%
%
%

\subsection{Representing integer numbers as matrices}

In the following, we use  $\mu^{p\times q}_n$ in Definition
\ref{DefNumericMatrixInterpReals} as a basis for the following representation 
mapping for \emph{integer numbers}.

\begin{definition}[Representing integer numbers as matrices]\label{DefNaturalNumbersAsMatrices}
Let $n\in\naturals$ be such that $n>1$ and 
$\mu_n$ be given as follows: for all $x\in\integers$, 
\[\mu_n(x)=\left \{
\begin{array}{rl}
\frac{x}{n}\cmatrix{n\times n}{1} & \mbox{if $n$ divides $x$}\\
xI_n & \mbox{otherwise}
\end{array}
\right .\]
We also define $\nu_n(x)=x\cmatrix{n}{1}$ to represent a number $x$ as a $n$-dimensional \emph{vector}.
\end{definition}
\begin{example}
The matrix 
$\mu_2(4)=
\left (
\begin{array}{cc}
2 & 2 \\
2 & 2
\end{array} \right )$
represents $4$ according to Definition \ref{DefNaturalNumbersAsMatrices}. 
\end{example}
Note that, for all $n\in\naturals$, $(\mu_n,\rho_n)$ 
(with $\rho_n$ as in Definition \ref{DefNumericMatrixInterpReals})
is a numeric matrix representation for integer numbers.
We obtain the following:



\begin{proposition}[Decrease of natural entries in matrices]\label{PropDecNaturalEntries}
Every matrix $A\in\naturals^{p\times q}$ such that $n=max(A)>1$ 
can be represented as a  matrix $A'\in\naturals^{np\times nq}$
such that, for all $m\in\positivenaturals$, 
$\rho_m(A)=\rho_{mn}(A')$ 
and $max(A)>max(A')$.
\end{proposition}
%
%
Obviously, Proposition \ref{PropDecNaturalEntries} entails the following.

\begin{corollary}[Natural matrices as bit matrices]\label{CoroNaturalMatricesAsBitMatrices}
Every matrix $A\in\naturals^{p\times q}$ 
can be represented as a bit matrix $A'\in\{0,1\}^{np\times nq}$ for some $n\in\positivenaturals$
and for all $m\in\positivenaturals$, $\rho_m(A)=\rho_{mn}(A')$.
\end{corollary}

\section{Representation of rational numbers below $1$}\label{SecRepOfRationalNumbers}

In this section, we investigate matrix representations (over the naturals) which can be used to deal with
rational numbers $\frac{1}{q}$ for some $q\in\positivenaturals$.
A $q$-square matrix $A_{\frac{1}{q}}$ which is almost null except for a single entry of value $1$
can be used to represent $\frac{1}{q}$ because $\rho_q(A_{\frac{1}{q}})=\frac{1}{q}$.
By Corollary \ref{CoroProductOfRepMappingInSquareMatrices}, for $A_p=p I_q$ we get 
$\rho_q(A_pA_{\frac{1}{q}})=\rho_q(A_{\frac{1}{q}}A_p)=\rho_q(A_p)\rho_q(A_{\frac{1}{q}})=\frac{p}{q}$.
Therefore, we can represent a rational number $\frac{p}{q}$ as a $q$-square matrix with a single entry
of value $p$. However, we have to change the size of the matrix if a different number $\frac{p'}{q'}$
with $q\neq q'$ is considered. 

\begin{remark}\label{RemRepOfRationalsAsMatricesOverNaturals}
Note that \emph{there is no generic representation  
of all rational numbers by using matrices over the naturals of a given dimension which is 
able to represent their values and appropriate comparisons among them}. 
Such a representation should be able to represent 
a decreasing sequence 
$1>\frac{1}{2}>\frac{1}{3}>\cdots>\frac{1}{n}>\cdots$ by
means of matrices $A_\frac{1}{n}\in\naturals^{p\times q}$ for all $n\in\positivenaturals$
satisfying $\rho_m(A_{\frac{1}{n}})>\rho_m(A_{\frac{1}{n+1}})$ for all 
$n\in\positivenaturals$. Equivalently, we should have 
$m\rho_m(A_{\frac{1}{n}})>m\rho_m(A_{\frac{1}{n+1}})$ for all 
$n\in\positivenaturals$. Since $m\rho_m(A_{\frac{1}{n}}),m\rho_m(A_{\frac{1}{n+1}})\in\naturals$, this would imply the existence of an infinite decreasing sequence of 
\emph{natural} numbers, which is not possible.
\end{remark}
Furthermore, the product of rational numbers $\frac{p}{q}$ and $\frac{p'}{q}$ represented as the $q$-square matrices indicated
above is \emph{not} preserved by the matrix product. Thus, in the following, we look for better representations.

\subsection{Use of nilpotent matrices}

\emph{Nilpotent} matrices are $n$-square matrices $B$ satisfying $B^k=\cmatrix{n\times n}{0}$ 
for some positive integer $k\in\positivenaturals$ 
\cite[Section 4.1]{Zhang_MatrixTheory_1999} 
(which is called the \emph{degree  of nilpotency} 
\cite[page 12]{LanTis_TheoryOfMatrices_1985}
or the \emph{index} of nilpotency of $B$ \cite[page 396]{Meyer_MatrixAnalysisAppliedAlgebra_2000}).
The degree of nilpotency $k$ of a $n$-square matrix $A$ is bounded by $n$:
$k\leq n$ \cite[Exercise 7.7.1]{Meyer_MatrixAnalysisAppliedAlgebra_2000}.
Given $n$, the following $n$-square matrix
(called a \emph{Jordan block}
\cite[Page 579]{Meyer_MatrixAnalysisAppliedAlgebra_2000}):
%
\[J_n = \left (
\begin{array}{c@{\hspace{0.3cm}}c@{\hspace{0.3cm}}c@{\hspace{0.3cm}}c}
0         & 1 & \\
           & \ddots & \ddots \\
           &             & \ddots & 1\\
           &             &             & 0
\end{array}
\right )
\]
(i.e., there are $n-1$ ones in the superdiagonal of $J_n$ and all other entries in
$J_n$ are zeroes) is nilpotent of degree $n$, i.e., $J^n_n=\cmatrix{n\times n}{0}$.

Write $J_n=[0,Z_1,\ldots,Z_{n-1}]$, where $Z_i$ is an almost null vector containing a single $1$ in
the $i$-th entry. For instance, $Z_1=(1,0,\ldots,0)^T$.
Then, it is not difficult to see that $J^p_n$ is obtained by introducing $p-1$ columns of zeros 
from the left side of the matrix and shifting columns $Z_i$ to the right just
throwing those which exceed the $n$-th position, i.e, $J^p_n=[0,\ldots,0,Z_1,\ldots,Z_{n-p}]$.

\begin{remark}\label{RemShiftMatrices}
$J_n$ Is also known as a \emph{shift} matrix because, for an arbitrary matrix $A$, 
$J_nA$ is obtained by shifting the rows of $A$ upwards by one position and introducing a new bottom 
row of zeroes.
Similarly $AJ_n$ shifts the columns of $A$ to the right and introduces a new leftmost column of zeroes.
\end{remark}
Note that, for all $p\in\naturals$,
\[\rho_m(J^p_n)=\left \{
\begin{array}{ll}
\frac{n-p}{m} & \mbox{ if } p<n\\
0 & \mbox{ if }p\geq n
\end{array}
\right .\]
The following result is obvious.

\begin{proposition}\label{PropReductiveJordanMatrices}
Let $m,n\in\positivenaturals$ and
 $0\leq p<q\leq n$.
Then, $\rho_m(J^p_n)=\rho_m((J^T_n)^p)>\rho_m((J^T_n)^q)=\rho_m(J^q_n)$.
For all $A\in\reals^{n\times r}$, $\rho_m(J^p_nA)\geq\rho_m(J^q_nA)$.
For all $A\in\reals^{r\times n}$, $\rho_m(AJ^p_n)\geq\rho_m(AJ^q_n)$.
\end{proposition}
In general  it is \emph{not} true that $\rho_m(J^pA)=\rho_m(AJ^p)$ for square matrices $A$.
%
%
Due to Proposition \ref{PropReductiveJordanMatrices} and Corollary \ref{CoroAdditionAndProductOfRepMapping}, 
for additive combinations $A$ of \emph{constant} and \emph{scalar} matrices we 
have the following:
\begin{corollary}
Let $m,n\in\positivenaturals$ and
$0\leq p<q\leq n$.
If $A\in\reals^{n\times r}$ (resp.  $A\in\reals^{r\times n}$) is an additive
combination of constant and scalar matrices, then 
$\rho_m(J^p_nA)=\rho_m(J^p_n)\rho_m(A)>\rho_m(J^q_n)\rho_m(A)=\rho_m(J^q_nA)$
(resp. $\rho_m(AJ^p_n)=\rho_m(A)\rho_m(J^p_n)>\rho_m(A)\rho_m(J^q_n)=\rho_m(AJ^q_n)$).
\end{corollary}
Thus, the following property of rational numbers $r=\frac{1}{n}$ for some $n\in\positivenaturals$
is simulated by the representation of integer numbers as (additive combinations of) 
constant or scalar matrices, and rational numbers $\frac{1}{n}$ as powers of Jordan blocks: 
for all $n\in\naturals$ and
positive rational number $r$, $0<r<1$, we have $n>nr>nr^2>\cdots$. 
%
%
\begin{example}\label{ExEncodingOfAHalf}
We can use $J_2$ to represent $\frac{1}{2}$: $\rho_2(J_2)=\frac{1}{2}$.
However, $J^2_2$, which is expected to correspond to $\frac{1}{4}$ does \emph{not} fit
this expectation: $\rho_2(J^2_2)=\rho_2(\cmatrix{2}{0})=0$.
\end{example}
%

\begin{theorem}\label{TheoProductOfScalarConstantAndJordanBlocksMatrices}
If $A_1,\cdots,A_N$ are $n$-square matrices such that, for all $i$, $1\leq i\leq N$,
either 
\begin{enumerate}
\item $A_i$ is scalar or constant, 
or 
\item $A_i=J^{p_i}_n$ for some $p_i\in\naturals$ and then both $A_{i-1}$ 
(if $i>1$) and $A_{i+1}$ (if $i<N$) 
are constant matrices, 
\end{enumerate}
then $\rho_n(\prod_{i=1}^NA_i)=\prod_{i=1}^N\rho_n(A_i)$.
\end{theorem}
As remarked above, it is \emph{not} possible to use matrices of natural numbers of a fixed size 
$n$ to represent \emph{all} fractions $\frac{1}{q}$ for $q\in\positivenaturals$.
Instead, we will consider the problem of representing \emph{finite subsets} 
$\cQ\subseteq\{\frac{1}{q}\mid q\in\positivenaturals\}$ by using $n$-\emph{square} (nilpotent) matrices
in such a way that the following property is fulfilled by the representation $(\mu,\rho)$: for all
$x,y\in\cQ$ such that $xy\in\cQ$, $\rho(\mu(x)\mu(y))=xy=\rho(\mu(x))\rho(\mu(y))$, 
i.e., the number $\rho(\mu(x)\mu(y))$ which corresponds to the matrix product $\mu(x)\mu(y)$ of matrices $\mu(x)$ and $\mu(y)$ 
representing $x\in\cQ$ and $y\in\cQ$ is exactly $xy\in\cQ$.
In Section \ref{SecTransfConstraints} we discuss how to take benefit from this.

The dimension $n$ of the matrices involved in the representation of $\cQ$ is determined by
the \emph{least} element in $\cQ$.
For instance, we can fix $n$ such that $\frac{1}{n}$ is the least number in $\cQ$.
Then,  an obvious representative for $\frac{1}{n}$ is $J_n^{n-1}$ because
$\rho_n(J_n^{n-1})=\frac{1}{n}$ (see Example \ref{ExEncodingOfAHalf}).
However, the feasibility of this simple approach depends on the other values in $\cQ$.

\begin{example}\label{ExRationalsAsMatrices1}
Let $\cQ=\{\frac{1}{2},\frac{1}{4}\}$. 
If we fix $J_4^3$ to be the representation of $\frac{1}{4}$, then the representation of $\frac{1}{2}$ 
should be $J^2_4$ (because $\rho_4(J^2_4)=\frac{1}{2}$). 
However, $\rho_4((J_4^2)^2)$ is \emph{not} $\frac{1}{4}$ as one could expect;
instead, $\rho_4((J_4^2)^2)=0$.
The following \emph{block} matrices of size $4$ whose blocks are combinations of 
(transposed) Jordan blocks can be used to represent $\cQ$ as required:
\[
\begin{array}{rcl@{\hspace{0.7cm}}rcl@{\hspace{0.7cm}}rcl}
Q_{\frac{1}{2}}& = &
\left (
\begin{array}{cc}
J_2 & J_2^T \\
\cmatrix{2\times 2}{0} & \cmatrix{2\times 2}{0}
\end{array} \right )
&
Q_{\frac{1}{4}} & = &
\left (
\begin{array}{cc}
\cmatrix{2\times 2}{0} & J_2J_2^T \\
\cmatrix{2\times 2}{0} & \cmatrix{2\times 2}{0}
\end{array} \right )
\end{array}
\]
Note that 
$\rho_4(Q_{\frac{1}{2}})=\frac{1+1}{4}=\frac{1}{2}$, 
$\rho_4(Q_{\frac{1}{4}})=\frac{0+1}{4}=\frac{1}{4}$, and
%
$Q_{\frac{1}{2}}Q_{\frac{1}{2}}=Q_{\frac{1}{4}}$.
\end{example}

\begin{example}\label{ExRationalsAsMatrices2}
Let $\cQ=\{\frac{1}{2},\frac{1}{4},\frac{1}{8}\}$. 
The following matrices of size $8$:
\[
\begin{array}{rcl@{\hspace{0.5cm}}rcl@{\hspace{0.5cm}}rcl}
Q_{\frac{1}{2}}
& = &
\left (
\begin{array}{cc}
J_4 & J_4^3 \\
\cmatrix{4\times 4}{0} & \cmatrix{4\times 4}{0}
\end{array} \right )
&
Q_{\frac{1}{4}}
& = &
\left (
\begin{array}{cc}
J^2_4 & \cmatrix{4\times 4}{0} \\
\cmatrix{4\times 4}{0} & \cmatrix{4\times 4}{0}
\end{array} \right )
\\[0.5cm]
Q_{\frac{1}{8}} 
& = &
\left (
\begin{array}{cc}
J^3_4 & \cmatrix{4\times 4}{0} \\
\cmatrix{4\times 4}{0} & \cmatrix{4\times 4}{0}
\end{array} \right )
\end{array}
\]
can be used to represent $\cQ$. Note that 
$\rho_8(Q_{\frac{1}{2}})=\frac{3+1}{8}=\frac{1}{2}$,
$\rho_8(Q_{\frac{1}{4}})=\frac{2+0}{8}=\frac{1}{4}$, 
$\rho_8(Q_{\frac{1}{8}})=\frac{1+0}{8}=\frac{1}{8}$,
%
$Q_{\frac{1}{2}}Q_{\frac{1}{2}}=Q_{\frac{1}{4}}$
and $Q_{\frac{1}{2}}Q_{\frac{1}{4}}=Q_{\frac{1}{4}}Q_{\frac{1}{2}}=Q_{\frac{1}{8}}$, as required.
\end{example}

\begin{example}\label{ExRationalsAsMatrices3}
Let $\cQ=\{\frac{1}{2},\frac{1}{3},\frac{1}{6}\}$. 
The following (block) matrices of size $6$ can be used to represent $\cQ$:
\[
\begin{array}{rcl@{\hspace{0.5cm}}rcl@{\hspace{0.5cm}}rcl}
Q_{\frac{1}{2}} 
& = &
\left (
\begin{array}{cc}
J_3 & (J^2_3)^T \\
\cmatrix{3\times 3}{0} & \cmatrix{3\times 3}{0}
\end{array} \right )
&
Q_{\frac{1}{3}} 
& = &
\left (
\begin{array}{cc}
J^2_3 & J_3(J^2_3)^T \\
\cmatrix{3\times 3}{0} & \cmatrix{3\times 3}{0}
\end{array} \right )
\\[0.5cm]
Q_{\frac{1}{6}} 
& = &
\left (
\begin{array}{cc}
\cmatrix{3\times 3}{0} & J^2_3(J^2_3)^T \\
\cmatrix{3\times 3}{0} & \cmatrix{3\times 3}{0}
\end{array} \right )
\end{array}
\]
Note that 
$\rho_6(Q_{\frac{1}{2}})=\frac{2+1}{6}=\frac{1}{2}$,
$\rho_6(Q_{\frac{1}{3}})=\frac{1+1}{6}=\frac{1}{3}$, and
$\rho_6(Q_{\frac{1}{6}})=\frac{0+1}{6}=\frac{1}{6}$.
Again, it is not difficult to see that 
$Q_{\frac{1}{2}}Q_{\frac{1}{3}}=Q_{\frac{1}{3}}Q_{\frac{1}{2}}=Q_{\frac{1}{6}}$.
Note, however, that if we add $\frac{1}{4}$ to $\cQ$, then we could \emph{not} use these matrices
for representing $\frac{1}{4}$.
In particular, $Q^2_{\frac{1}{2}}=Q_{\frac{1}{3}}$, i.e., $\rho_6(Q^2_{\frac{1}{2}})\neq\frac{1}{4}$
as should be the case.
\end{example}
%

%
%

\section{Matrix interpretations revisited}\label{SecMatrixBlockBasedMatrixInterp}

As remarked above, termination problems in term rewriting are usually translated into 
conjunctions of \emph{weak} or \emph{strict}
\emph{symbolic constraints} like  $s\succeq t$ or $s\succ t$
between terms $s$ and $t$
coming from (parts of) the TRS.
In order to check the satisfaction of these constraints, we need to use \emph{term (quasi-)orderings}.
Such term orderings can be obtained by giving appropriate \emph{interpretations} 
to the function symbols of a signature. 
Given a signature $\Symbols$, an
$\Symbols$-algebra is a pair $\cA=(\SemDomain,\Symbols_\SemDomain)$, where 
$\SemDomain $ is a set 
and $\Symbols_\SemDomain $ is a set of mappings $f_\cA: \SemDomain^k\to \SemDomain$ for each 
$f\in\Symbols$ where $k=ar(f)$. 
For a given valuation mapping $\alpha:\Variables\to\SemDomain$, the evaluation mapping 
$[\alpha]:\Terms\to \SemDomain$ is inductively defined by 
$[\alpha](x)=\alpha(x)$ if $x\in\Variables$ and 
$[\alpha](f(t_1,\ldots,t_k))=f_\cA([\alpha](t_1),\ldots,[\alpha](t_k))$ 
for $x\in\Variables$, $f\in\Symbols$, $t_1,\ldots,t_k\in\Terms$. 
Given a term $t$ with $\var(t)=\{x_1,\ldots,x_n\}$, we write 
$[t]$ to denote the {\em function} $F_t: \SemDomain^n\to \SemDomain$ given 
by $F_t(a_1,\ldots,a_n)=[\alpha_{(a_1,\ldots,a_n)}](t)$ for each 
tuple $(a_1,\ldots,a_n)\in \SemDomain^n$, where $\alpha_{(a_1,\ldots,a_n)}(x_i)=a_i$ for 
$1\leq i\leq n$.
%
We can define a stable quasi-ordering $\succeq$ on terms given by 
$t\succeq s$ if and only  if
$[\alpha](t)\succeq_\SemDomain[\alpha](s)$, for all $\alpha:\Variables\to\SemDomain $, where $\succeq_\SemDomain $ is a  
quasi-ordering on $\SemDomain$.
We can define a stable strict ordering $\sqsupset$ on terms by
$t \sqsupset s$ if 
$[\alpha](t)\succ_\SemDomain[\alpha](s)$, for all $\alpha:\Variables\to \SemDomain$,
where $\succ_ \SemDomain $ is a strict ordering on $\SemDomain$.
%
%


A matrix interpretation for a $k$-ary symbol $f$ is a  linear expression $F_1x_1+\cdots+F_kx_k+F_0$
where the $F_1,\ldots, F_k$ are (square) matrices
of $n\times n$ natural 
numbers and the variables $x_1,\ldots,x_k$ (and also the constant term $F_0$) are $n$-tuples of natural numbers 
\cite{EndWalZan_MatrixInterpretations_JAR08}.
An expression like $F\vec{x}$, where $F$ is an $n$-square matrix and 
$\vec{x}$ is an $n$-tuple of numbers,
is interpreted as the usual matrix-vector product, i.e., the $i$-th
component $y_i$ of $\vec{y}=F\vec{x}$ is $y_i=\sum^n_{j=1}F_{ij}x_j$.
Matrices and vectors are compared by using an \emph{entrywise} ordering: for 
$A,B\in\reals^{p\times q}$, we write $A\geq B$ if $A_{ij}\geq B_{ij}$ for all $1\leq i\leq p$, $1\leq j\leq q$
\cite[Chapter 15]{LanTis_TheoryOfMatrices_1985}.
We also write $A>B$ if $A\geq B$ and $A_{11}>B_{11}$ \cite{EndWalZan_MatrixInterpretations_JAR08}.
Note that this also describes how to compare tuples of numbers.
In \cite{AlaLucNav_ProvingTerminationMatrixInterpretationsOverReals_WST09,%
AlaLucNav_UsingMatrixInterpretationsOverReals_PROLE09}, Endrullis et al.'s approach was 
extended to matrix interpretations with real numbers in matrix entries.

Here, we generalize the notion of matrix interpretation in the following ways:
\begin{enumerate}
\item We consider a domain $\blockvectors{n}{b}{N}$ of \emph{block-based} tuples \emph{over $N$}, 
i.e., $n$-tuples consisting of 
$\beta=\frac{n}{b}$ tuples of size $b\in\positivenaturals$ which are \emph{constant} tuples 
$c\cmatrix{b}{1}$ for some number $c\in N$.
If we take $b=1$ and $N=\naturals$ or $N=\nonnegativereals$, then we are in the original approaches
\cite{EndWalZan_MatrixInterpretations_JAR08} and \cite{AlaLucNav_ProvingTerminationMatrixInterpretationsOverReals_WST09,%
AlaLucNav_UsingMatrixInterpretationsOverReals_PROLE09}, respectively.
Note, however, that given $n\in\positivenaturals$, only divisors $b$ of $n$ can be used to
establish blocks within $n$-tuples of numbers. 
\item Matrices $F$ in matrix expressions
interpreting symbols $f\in\Symbols$ will be \emph{block
matrices}
{\footnotesize $\left (
\begin{array}{ccc}
F_{11} & \cdots & F_{1\beta}\\
\vdots & \ddots & \vdots\\
F_{\beta 1} & \cdots & F_{\beta \beta}
\end{array}
\right )
$}%
such that $F_{ij}=C_{ij}+S_{ij}$ is a sum  of a $b$-square \emph{constant} matrix 
$C_{ij}=c_{ij}\cmatrix{b\times b}{1}$, where $c_{ij}\in N$, and 
a $b$-square \emph{scalar} matrix  $S_{ij}=s_{ij}I_b$, where $s_{ij}\in N$, for all $1\leq i,j\leq \beta$.
This is necessary for soundness of the obtained algebraic interpretation: the product
of one of such matrices by a block-based tuple as above produces a block-based tuple as above, i.e.,
$F\vec{v}\in\blockvectors{n}{b}{N}$ for all $\vec{v}\in\blockvectors{n}{b}{N}$.
Furthermore, such matrices are \emph{closed} under addition and matrix product.
This is essential to obtain matrices of this kind during the interpretation of the terms, where nested
symbols yield products and sums of matrices after interpreting them.
Again, if $b=1$, we are in the usual case for matrix interpretations. 
\item  Given a matrix valuation $\rho$, and matrices $A,B$, we let
$A\geq_\rho B$ if $\rho(A) \geq_N\rho(B)$, where $\geq_N$ is an ordering over $N$.
%
%
Given $\delta>0$, the following ordering over (real) numbers is used \cite{Lucas_PolOverRealsTheoPrac_TIA05}:
$x>_\delta y$ if $x-y\geq\delta$.
If $\delta=1$ and $x,y$ are natural numbers, we obtain the usual well-founded
ordering among natural numbers $>_\naturals$.
Now, the following (strict and well-founded) ordering $>_{\rho,\delta}$ 
(or just $>_\rho$ or even $>$ if it is
clear from the context) on $n$-tuples of \emph{nonnegative} numbers is considered:
$\vec{x}>_{\rho,\delta}\vec{y}$ if $\rho(x)>_\delta\rho(y)$.
Clearly, $(\blockvectors{n}{b}{\naturals},>_{\rho,1})$ and
$(\blockvectors{n}{b}{\nonnegativereals},>_{\rho,\delta})$ are well-founded orderings.

The previous orderings do \emph{not} take into account the block structure of matrices or
tuples. The following one does: for matrices $A,B$ with blocks $A_{ij}$ and $B_{ij}$
for $1\leq i,j\leq \beta$, we write 
$A\geq_{\rho}^bB$ if 
$A_{ij}\geq_{\rho}B_{ij}$ for 
all $1\leq i,j\leq \beta$.
Similarly, $A>_{\rho,\delta}^bB$ if $A_{11}>_{\rho,\delta}B_{11}$ and $A_{ij}\geq_{\rho}B_{ij}$ for all $1\leq i,j\leq \beta$. 
These definitions are adapted to tuples in the obvious way.
\end{enumerate}

\begin{definition}[Block-based matrix interpretation]\label{DefBlockBasedMatrixInterp}
Let $\Symbols$ be a signature, $n,b\in\positivenaturals$ be such that $b$ divides $n$,
and $\beta=\frac{n}{b}$. 
An \emph{(n,b)-block-based matrix interpretation} is  an 
$\Symbols$-algebra $\cA=(A,\Symbols_\cA)$ such that
\begin{enumerate}
\item $A=\blockvectors{n}{b}{N}$ for $N=\naturals$ or $N=\nonnegativereals$, and
\item $\Symbols_\cA$ consists of matrix functions $[f](x_1,\ldots,x_k)=F_1x_1+\cdots+F_kx_k+F_0$
which are closed in $A$ and where, for all $f\in\Symbols$, $1\leq i\leq k$,
\begin{enumerate}
\item $F_i$ is an $n$-square  matrix of $\beta\times\beta$ blocks of $b$-square matrices 
$C_{j\ell}+S_{j\ell}$
such that $C_{j\ell}$ is a constant matrix, and $S_{j\ell}$ is a scalar matrix for all $1\leq j\leq\beta$
and $1\leq \ell\leq\beta$. 
\item $F_0=(\vec{c}_1^T\cdots\vec{c}_\beta^T)^T$  consists of 
$\beta$ $b$-tuples
$\vec{c}_j=c_j\cmatrix{b}{1}$ for some $c_j\in N$.
\end{enumerate}
\end{enumerate}
We make the orderings which are going to be used explicit by adding them to the
pair $(A,\Symbols_\cA)$, thus specifying an \emph{ordered} algebra: 
$(A,\Symbols_\cA,\geq_{\rho_n})$, $(A,\Symbols_\cA,\geq_{\rho_b}^b)$, etc.
\end{definition}

\section{Solving and transforming matrix constraints}\label{SecTransfConstraints}

In the following, we consider two kinds of constraint solving problems which are relevant in
proofs of termination.

\subsection{Testing universally quantified symbolic constraints}

Proofs of termination in term rewriting involve solving \emph{weak} or \emph{strict}
symbolic constraints $s\succeq t$ or $s\sqsupset t$ between terms $s$ and $t$ coming from (parts of) the rules of the TRS
where the variables in $s$ and $t$ are universally quantified in the corresponding constraint.
Here, $\succeq$ and $\sqsupset$ are (quasi-)orderings on terms satisfying appropriate conditions
\cite{ArtGie_TermOfTermRewUsingDepPairs_TCS00,%
BorRub_OrderingsAndConstraintsForTermination_RCP07,%
Dersh_TerminationRewriting_JSC87}. 
\begin{example}\label{ExSymbolicCSofMainExample}
The following symbolic constraints must be checked in order to guarantee termination of the
TRS $\cR$ in Example \ref{ExPaperRationalMatrices}:
%
\begin{eqnarray}
\forall X (f(f(X)) & \succeq & f(g(f(X))))\label{CSEmbeddingRule}\\
\forall X (f(g(f(X))) & \succeq & X)\label{CSProjectionRule}\\
\forall X (F(f(X)) & \sqsupset & F(g(f(X))))\label{CSEmbeddingDP1}\\
\forall X (F(f(X)) & \sqsupset & F(X))\label{CSEmbeddingDP2}
\end{eqnarray}
Here, variables $X$ range on terms in $\Terms$, 
$\succeq$ and $\sqsupset$
are intended to be interpreted as a monotonic and stable quasiordering on terms, and
a well-founded and stable ordering on terms, respectively.
\end{example}
In the so-called \emph{interpretation method} for checking symbolic constraints as the ones
in Example \ref{ExSymbolicCSofMainExample}, we use appropriate 
\emph{ordered $\Symbols$-algebras} to \emph{generate} the necessary orderings 
(see Section \ref{SecMatrixBlockBasedMatrixInterp}).
In our setting, we are interested in investigating the use of matrix algebras.
The following result shows how to \emph{decrease} the value of some of the entries in a  
block-based matrix algebra over the naturals to obtain an \emph{equivalent} 
block-based matrix algebra which uses bigger matrices.

\begin{theorem}\label{TheoTransfOfMatrixInterpOverTheNaturals}
Let $\Symbols$ be a signature and $\genconstraint$  be a symbolic constraint $\forall x(s \compOp t)$
for terms $s,t\in\Terms$.
Let $\cA=(\blockvectors{n}{b}{\naturals},\Symbols_\cA,\geq^b_{\rho_b})$ be an $(n,b)$-block-based matrix interpretation 
over the naturals and $N$ be the 
maximum of all entries occurring in any matrix or vector in $\cA$.
Let $\cB=(B,\Symbols_{\cB},\geq^{bN}_{\rho_{bN}})$ be an $(Nn,Nb)$-block-based matrix interpretation where
$B=\blockvectors{Nn}{Nb}{\naturals}$ and for all $f\in\Symbols$,
$[f]_\cB(x_1,\ldots,x_k)=\mu_N(F_1)x_1+\cdots+\mu_N(F_k)x_k+\nu_N(F_0)$ iff
$[f]_\cA(x_1,\ldots,x_k)=F_1x_1+\cdots+F_kx_k+F_0$.
Then, $\cA$ satisfies $\genconstraint$ if and only if $\cB$ satisfies $\genconstraint$.
\end{theorem}

\begin{example}
Consider the following TRSs \cite[Example 3]{EndWalZan_MatrixInterpretations_JAR08}: 
\begin{eqnarray}
\cR: f(a,g(y),z) & \to &  f (a,y,g(y))\label{Ex3_EWZ08_Rrule1}\\
f(b,g(y),z) & \to & f(a,y,z)\label{Ex3_EWZ08_Rrule2}\\
a & \to & b\label{Ex3_EWZ08_Rrule2}\\
\cS : f(x,y,z) & \to &  f(x,y,g(z))\label{Ex3_EWZ08_Srule1}
\end{eqnarray}
In order to prove termination of $\cR$ \emph{relative} to $\cS$ (written $SN(R/S)$), we have to
check that $\forall x(l\succeq r)$ holds for all rules $l\to r\in\cR\cup\cS$.
Endrullis et al.\ use the following matrix interpretation for that:
\[\begin{array}{rcl@{\hspace{0.5cm}}rcl@{\hspace{0.5cm}}rcl@{\hspace{0.5cm}}rcl@{\hspace{0.5cm}}rcl}
{}[a] & = & {} 
\left (
\begin{array}{cc}
1\\
0
\end{array} \right )
\\[0.5cm]
{}[b] & = & \left (
\begin{array}{c}
0\\
0
\end{array} \right ) 
\\[0.5cm]
{}[f](x,y,z) & = & {} 
\left (
\begin{array}{cc}
1 & 0 \\
0 & 0
\end{array} \right ) 
x + 
\left (
\begin{array}{cc}
1 & 2 \\
0 & 0
\end{array} \right ) 
y + 
\left (
\begin{array}{cc}
1 & 0 \\
0 & 0
\end{array} \right ) 
z + 
\left (
\begin{array}{c}
0\\
0
\end{array} \right ) 
\\[0.5cm]
{}[g](x) & = & {} \left (
\begin{array}{cc}
1 & 0 \\
1 & 1
\end{array} \right ) 
x + 
\left (
\begin{array}{cc}
0 \\
1
\end{array} \right )
\end{array}
\]
By using Theorem \ref{TheoTransfOfMatrixInterpOverTheNaturals}, we 
conclude that the following block-based matrix interpretation
with bit matrices of dimension 4 can also do the work.

{\footnotesize
\[\begin{array}{rcl@{\hspace{0.5cm}}rcl@{\hspace{0.5cm}}rcl@{\hspace{0.5cm}}rcl@{\hspace{0.5cm}}rcl}
{}[a] & = & {} 
\left (
\begin{array}{cc}
\vec{1}_2\\
\vec{0}_2
\end{array} \right )
\\[0.5cm]
{}[f](x,y,z) & = & {} 
\left (
\begin{array}{cc}
I_2 & \cmatrix{2\times 2}{0} \\
\cmatrix{2\times 2}{0} &  \cmatrix{2\times 2}{0}
\end{array} \right ) 
x + 
\left (
\begin{array}{cc}
I_2 & \cmatrix{2\times 2}{1} \\
\cmatrix{2\times 2}{0} & \cmatrix{2\times 2}{0}
\end{array} \right ) 
y + 
\left (
\begin{array}{cc}
I_2 & \cmatrix{2\times 2}{0}\\
\cmatrix{2\times 2}{0} & \cmatrix{2\times 2}{0}
\end{array} \right ) 
z + 
\left (
\begin{array}{c}
\cmatrix{2}{0}\\
\cmatrix{2}{0}
\end{array} \right ) 
\\[0.5cm]
{}[b] & = & \left (
\begin{array}{c}
\cmatrix{2}{0}\\
\cmatrix{2}{0}
\end{array} \right ) 
\\[0.5cm]
{}[g](x) & = & {} \left (
\begin{array}{cc}
I_2 & \cmatrix{2\times 2}{0} \\
I_2 & I_2
\end{array} \right ) 
x + 
\left (
\begin{array}{cc}
\cmatrix{2}{0} \\
\cmatrix{2}{1}
\end{array} \right )
\end{array}
\]}
\end{example}
%

\subsection{Solving existentially quantified arithmetic constraints}

In many applications, when algebraic interpretations have to be \emph{generated}
rather than given by the user, it is usual to work with \emph{parametric} interpretations.

\begin{example}\label{ExParametricInterpAndConstraints}
For the symbols occurring in $\cR$ in Example \ref{ExPaperRationalMatrices}, we can 
consider the following linear \emph{parametric} interpretation:
%
\[\begin{array}{rcl@{\hspace{1cm}}rcl@{\hspace{1cm}}rcl}
{}[f](x) & = & f_1x + f_0 &
{}[g](x) & = & g_1 x + g_0 &
{}[F](x) & = & F_1 x + F_0
\end{array}
\]
where $f_1,g_1$, and $F_1$ are expected to be $n$-square matrices for some $n\in\positivenaturals$
(the case $n=1$ corresponds to a linear polynomial intepretation) 
and $f_0,g_0,F_0$ are $n$-tuples.
The satisfaction of the following \emph{arithmetic} constraints\footnote{By lack of space,
we cannot explain how these constraints are obtained. Full details about this standard procedures
can be found elsewhere, see, e.g., \cite{ConMarTomUrb_MechProvTermPolInterp_JAR06,%
EndWalZan_MatrixInterpretations_JAR08,Lucas_PolOverRealsTheoPrac_TIA05,%
Lucas_PracticalUseOfPolyOverReals_PPDP07}.}
%
\begin{eqnarray}
f_1f_1 & \geq & f_1g_1f_1\label{CSexampleRule1variable}\\
f_1f_0 + f_0 & \geq & f_1g_1f_0+f_1g_0+ f_0\label{CSexampleRule1constant}\\
f_1g_1f_1 & \geq & 1\label{CSexampleRule2variable}\\
f_1g_1f_0+f_1g_0+f_0 & \geq  & 0\label{CSexampleRule2constant}\\
F_1f_1 & \geq & F_1g_1f_1\label{CSexampleDP1variable}\\
F_1f_0 + F_0 & > & F_1g_1f_0+F_1g_0+ F_0 \label{CSexampleDP1constant}\\
F_1f_1& \geq & F_1\label{CSexampleDP2variable}\\
F_1f_0 + F_0 & > & 0\label{CSexampleDP2constant}
\end{eqnarray}
is necessary to ensure that $\cR$ is terminating.
\end{example}
By a parametric matrix interpretation we mean a matrix intepretation where the entries in
matrices are not numbers but rather \emph{parametric coefficients}, i.e., variables for which
we have to provide a numeric value, depending on some (existential) constraints.

In general, given a set of variables $\Variables$, we consider here symbolic arithmetic constraints 
of the form 
$s \compOp t$ for $\compOp\:\in\{\geq,>\}$, where
$s$ is of the form $\sum_{i=1}^{m_s}s_i$ for 
$m_s>0$ and 
$s_i=s_{i1}\cdots s_{im_{s,i}}$ with $m_{s,i}>0$ for all  $1\leq i\leq\sigma_s$
and $s_{ij}\in\Variables$ ($t$ would have an analogous structure).
Furthermore, we assume existential quantification over all variables occurring in
$s$ and $t$.
Note in Example \ref{ExParametricInterpAndConstraints} that we use \emph{constants} 
like $0$ and $1$. They are handled as `special'
variables which will receive the intended interpretation.

Consider a \emph{valuation} 
$\eta:\Variables\to N$ 
for the variables in $\Variables$ as numbers in $N$.
%
%
Here, we consider $N=\naturals\cup\cQ$ for some finite subset of rational numbers 
$\cQ\subseteq\rationals-\naturals$ satisfying some conditions.
We are interested in representing numbers $\eta(x)$ as matrices $\mu(\eta(x))$
as discussed above.
Note that we \emph{cannot} represent arbitrary rational numbers by using matrices over the
naturals (Remark \ref{RemRepOfRationalsAsMatricesOverNaturals}).
Still, when dealing with finite sets $\cC$ of arithmetic restrictions, 
we can restrict the attention to those rational numbers which are required to 
check its satisfaction for a given valuation $\eta$.
This includes not only rational numbers $\eta(x)$ which are assigned to $x\in\Variables$,
but also those which could occur during the evaluation of an arithmetic expression
due to products $\eta(x)\eta(y)$ 
of rational numbers $\eta(x)$ and $\eta(y)$ which have been associated to variables $x$ and
$y$.
The idea is that $\cQ$ should contain such rational numbers.
\begin{definition}[Compatible domain of rational numbers]
Let $\eta:\Variables\to\naturals\cup\cQ$ be a valuation
for some $\cQ\subseteq\rationals-\naturals$ and $\cC$ be a set of arithmetic constraints. 
 Given a
multiplicative component $s=s_{1}\cdots s_{m}$ of an arithmetic expression in
a constraint in $\cC$, let $I_s=\{i_1,\ldots,i_k\}$ be the set of indices of variables in $s$ whose valuation 
is a rational (and noninteger) number, i.e., for all $i\in I_s$, $\eta(s_i)\in\rationals-\naturals$. 
We say that $\cQ$ is \emph{compatible} with $\cC$ and $\eta$ if 
$\prod_{j\in J}\eta(s_i)\in\cQ$ for all multiplicative components $s$ in $\cC$ and
$J\subseteq I_s$ such that $J\neq\emptyset$.
\end{definition}
\begin{example}\label{ExNumValuationOfMainExample}
The numeric valuation which corresponds to the poynomial interpretation in Example \ref{ExPaperRationalMatrices}  is:
\[\begin{array}{rcl@{\hspace{0.5cm}}rcl@{\hspace{0.5cm}}rcl@{\hspace{0.5cm}}rcl}
\eta(1) & = & 1 &
\eta(0) & = & 0 & 
\eta(f_1) & = & 2 &
\eta(f_0) & = & 2 \\
\eta(g_1) & = & \frac{1}{2} &
\eta(g_0) & = & \frac{1}{2} & 
\eta(F_1) & = & 1 &
\eta(F_0) & = & 0 
\end{array}\]
The set $\cQ=\{\frac{1}{2}\}$ is compatible with this valuation and with the constraints $\cC$
in Example \ref{ExParametricInterpAndConstraints}.
Consider $\cC'=\cC\cup\{f_1g_1f_1\geq f_1g_1f_1g_1\}$. Now, $\cQ$ is \emph{not} compatible
with $\cC'$ and $\eta$ because we have that $\eta(g_1)\eta(g_1)=\frac{1}{4}\not\in\cQ$.
If we add $\frac{1}{4}$ to $\cQ$, then we get compatibility with $\cC'$.
\end{example}
A valuation $\eta$ is extended to terms and constraints by $\eta(s\compOp t)=\eta(s)\compOp_\rationals\eta(s)$,
$\eta(s_1+\cdots+s_m)=\eta(s_1)+\cdots+\eta(s_m)$, and 
$\eta(x_1\cdots x_m)=\eta(x_1)\cdots\eta(x_m)$.

\begin{remark}
In general, we assume that $+$ is commutative, but (as happens with the matrix
product) we do \emph{not} assume commutativity of the product in arithmetic expressions
or their valuations.
\end{remark}
Now we have to extend $\mu_n$ in Definition \ref{DefNaturalNumbersAsMatrices} to deal with 
rational numbers in $\cQ$.
\begin{remark}
In contrast to natural numbers, we have no systematic way to associate matrices to rational
numbers yet.
In Section \ref{SecRepOfRationalNumbers} we have investigated some partial solutions to this problem.
In particular, Examples \ref{ExEncodingOfAHalf}, 
\ref{ExRationalsAsMatrices1}, \ref{ExRationalsAsMatrices2}, 
and \ref{ExRationalsAsMatrices3}, show that
the \emph{dimension} $n$ of the considered matrices and tuples
heavily depend on the numbers  in $\cQ$.
On the other hand, these examples also provide useful encodings for rational numbers 
which are frequently used in automatic proofs of termination with polynomial or matrix interpretations
\cite{BorrallerasEtAl_SolvingPolyAritithSMT_CADE09,FuhsEtAl_SearchTechRatPolyOrd_AISC08,%
Lucas_PracticalUseOfPolyOverReals_PPDP07}.
\end{remark}
Assume that $\mu_n$ has been extended to each $x\in\cQ$ in such a way that:
for all $x,y\in\cQ$ such that $xy\in\cQ$, $\rho_n(\mu_n(x) \mu_n(y))=xy=\rho_n(\mu_n(x))\rho_n(\mu_n(y))$.

In our setting, variables in $\Variables$ are interpreted not only as matrices but some of them as 
\emph{vectors}.
Assume that $\Variables_0\subseteq\Variables$ must be interpreted in this way and that the constraints
in $\cC$ are consistent with this, i.e., whenever a variable $x_0\in\Variables_0$ occurs in a constraint
$\genconstraint\in\cC$ of the form $s\compOp t$, each multiplicative term in $s$ and $t$ must contain 
a single variable $y_0\in\Variables_0$ which must be at the end of the term.
\begin{example}
For the constraints in Example \ref{ExParametricInterpAndConstraints}, we have 
$\Variables_0=\{f_0,F_0,g_0,0\}$.
\end{example}
The vectorial (or $n$-tuple) representation of  $x\in\naturals\cup\cQ$ by $\mu_n$ is $\mu_n(x)\cmatrix{n}{1}$.

\begin{example}\label{ExPaperRationalMatricesInduced}
The matrices over the naturals which correspond to $\eta$ and
$\cQ$ in Example \ref{ExNumValuationOfMainExample} is (with the encoding of
$\frac{1}{2}$ in Example \ref{ExEncodingOfAHalf}) are:
%
\[\begin{array}{rcl@{\hspace{0.4cm}}rcl@{\hspace{0.4cm}}rcl@{\hspace{0.5cm}}rcl}
\mu(\eta(f_1)) & = & 
\cmatrix{2\times 2}{1} 
= \left (
\begin{array}{cc}
1 & 1 \\
1 & 1
\end{array} \right ) 
& 
\mu(\eta(f_0)) & = & 2\:\cmatrix{2}{1}  = 
\left (
\begin{array}{c}
2\\
2
\end{array} \right ) 
\\[0.5cm]
\mu(\eta(g_1)) & = & 
J_2 = 
\left (
\begin{array}{cc}
0 & 1 \\
0 & 0
\end{array} \right ) 
&
\mu(\eta(g_0)) & = &
J_2\cmatrix{2}{1} =
\left (
\begin{array}{c}
1\\
0
\end{array}
\right )
\\[0.5cm]
\mu(\eta(F_1)) & = & 
I_2 = 
\left (
\begin{array}{cc}
1 & 0 \\
0 & 1
\end{array} \right ) &
\mu(\eta(F_0)) & = & 
0\:\cmatrix{2}{1} =
\left (
\begin{array}{c}
0\\
0
\end{array}
\right )
\\
\mu(\eta(1)) & = & 
I_2 = 
\left (
\begin{array}{cc}
1 & 0 \\
0 & 1
\end{array} \right ) &
\mu(\eta(0)) & = & 
0\:\cmatrix{2}{1} =
\left (
\begin{array}{c}
0\\
0
\end{array}
\right )\\
\end{array}
\]
Accordingly, the matrix interpretation over the naturals which corresponds to the polynomial interpretation 
over the rationals in Example \ref{ExPaperRationalMatrices} is:
%
\[\begin{array}{rcl@{\hspace{0.5cm}}rcl@{\hspace{0.5cm}}rcl}
{}[f](x) & = & {} \left (
\begin{array}{cc}
1 & 1 \\
1 & 1
\end{array} \right ) x + \left (
\begin{array}{c}
2\\
2
\end{array} \right ) &
{}[g](x) & = & {} \left (
\begin{array}{cc}
0 & 1 \\
0 & 0
\end{array} \right ) x 
+
\left (
\begin{array}{c}
1\\
0
\end{array}
\right )
\\[0.5cm] 
{}[F](x) & = & {} \left (
\begin{array}{cc}
1 & 0 \\
0 & 1
\end{array} \right ) x
\end{array}
\]
This matrix interpretation induced from the polynomial interpretation over the 
rationals in Example \ref{ExPaperRationalMatrices} can also be used to solve the constraints in the example.
\end{example}
Our technique could be used to translate matrix interpretations \emph{over the rationals} into matrix interpretations
\emph{over the naturals} by just applying the previous translation to the \emph{entries} of the matrix interpretation over the
rationals instead to the coefficients of the polynomial interpretation.


\section{Conclusions}\label{SecConclusions}

We have investigated matrix representations of natural and rational numbers which can be used
to simulate the arithmetics of natural and rational numbers, respectively.
We have introduced the notion of \emph{numeric matrix representation} 
(Definition \ref{DefNumMatrixRep}) which associates a number
to a matrix and viceversa and proved that, by using some specific representations
(Definitions \ref{DefNumericMatrixInterpReals} and \ref{DefNaturalNumbersAsMatrices}),
the arithmetic of natural numbers is preserved.
Furthermore, we have proved that every matrix interpretation over the naturals has an
associated bit
matrix interpretation of (usually) bigger size of the same associated value 
(Corollary \ref{CoroNaturalMatricesAsBitMatrices}).
We have investigated the representation of rational numbers by using matrices of natural numbers.
We have proved that this problem has no general solution but we have found some suitable
trade-offs for finite subsets of rational numbers by using nilpotent matrices consisting of Jordan blocks.
Then we have introduced the notion of block-based matrix interpretation (Definition \ref{DefBlockBasedMatrixInterp})
which generalizes existing approaches to matrix interpretations.
We use it to transform matrix interpretations over the naturals into matrix interpretations over $\{0,1\}$,
and also to transform matrix interpretations over the rationals into matrix
interpretations over the naturals.

The question posed in the introduction: 
\emph{are rational numbers somehow \emph{unnecessary} when dealing with matrix interpretations?} 
could not be answered in full generality due to the lack of a general procedure for building matrix
representations for arbitrary finite sets of rational numbers. Of course, this is a main topic of further
research and we think that we have settled a good starting point in considering the use of combinations of Jordan blocks as in Examples \ref{ExEncodingOfAHalf}, 
\ref{ExRationalsAsMatrices1}, \ref{ExRationalsAsMatrices2}, 
and \ref{ExRationalsAsMatrices3}.
Nevertheless, our results suggest that the use of matrices over the 
naturals of \emph{big size} can somehow play the role of rational numbers in interpretations
of smaller size, and in particular in linear polynomial intepretations over the rationals.
This does \emph{not} mean that implementing polynomial or matrix intepretations over the rationals is
not useful anymore and that natural numbers should be used everywhere.
In fact, working with matrix interpretations of big size is computationally expensive.
Another interesting consequence of our analysis is the connection between dimension of matrices 
over the naturals and value of their entries via Proposition  \ref{PropDecNaturalEntries} 
and Corollary \ref{CoroNaturalMatricesAsBitMatrices}.
Roughly speaking, these results can be interpreted by saying that bigger dimensions of
matrices permit the use of smaller entries. In practice, most tools put strong numeric 
bounds to the coefficients or entries of the interpretations.
Our results suggest that increasing such bounds could have a similar effect to increasing the
size of the matrices.
A more precise analysis about the trade-offs in design and efficiency which these considerations
could lead to is also subject for future work.



\end{document}